\documentclass[aps,pra,reprint,a4paper]{revtex4-1}
\usepackage{amsmath}
\usepackage{amssymb}
\usepackage{graphicx}
\usepackage{color}

\newcommand{\bra}[1]{\ensuremath{\left\langle #1 \right|}}
\newcommand{\ket}[1]{\ensuremath{\left| #1 \right\rangle}}
\newcommand{\braket}[2]{\ensuremath{\left\langle #1 | #2 \right\rangle}}
\newcommand{\figref}[1]{Fig. \ref{#1}}
\newcommand{\correction}[1]{{#1}}

\begin{document}
\title{Feedback-controlled adiabatic quantum computation}
\author{R.D. Wilson} 
\email{R.D.Wilson@lboro.ac.uk}
\author{A.M. Zagoskin} 
\author{S. Savel'ev}
\author{M.J. Everitt}
\affiliation{Department of Physics, Loughborough University, Loughborough, Leicestershire LE11 3TU, UK}
\author{Franco Nori}
\affiliation{Advanced Science Institute, RIKEN, Wako-shi, Saitama 351-0198, Japan}
\affiliation{Physics Department, The University of Michigan, Ann Arbor, MI 48109-1040, USA}
\date{\today}
\begin{abstract}
	We propose a simple feedback-control scheme for adiabatic quantum computation with superconducting flux qubits. The proposed method makes use of existing on-chip hardware to monitor the ground state curvature, which is then used to control the computation speed in order to maximise the success probability. We show that this scheme \correction{can provide a polynomial speed-up in performance} and that it is possible to choose a suitable set of \correction{ feedback-control} parameters for an arbitrary problem Hamiltonian.
\end{abstract}
\maketitle

\section{Introduction}

	Adiabatic quantum computing (AQC) is a promising paradigm of quantum computation. It was first proposed as a possible method of solving NP-complete problems in a more efficient manner than is possible classically \cite{Farhi2000}\correction{. It} has since been shown to be equivalent to other universal models of quantum computation \cite{Aharonov2004,Mizel2007}. There have also been some significant advances in the experimental implementation of AQC \cite{vanderPloeg2007,Johnson2011,Bian2012} using superconducting flux qubits \cite{You2011,Zagoskin2011}. 
	
	In AQC, the ground state of a Hamiltonian encoding the problem is reached by adiabatic evolution from a configuration with an easily-reachable ground state. The final ground state will encode the solution to the problem. This process can be described by a Hamiltonian of the form
\begin{equation}
	\mathcal{H}\left[\lambda(t)\right]=H_{p}+\lambda(t)H_b,
	\label{eqn:Hamiltonian}
\end{equation}
where $H_p$ describes the problem, $H_b$ describes a strong applied bias (such that $H_b \gg H_p$ and $\left[H_{b},H_{p}\right]\neq0$) and the bias strength, $\lambda(t)$, is slowly varied from $1$ to $0$. The instantaneous energy eigenvalues, $E_{\ell}\left(\lambda\right)$, and eigenstates, $\ket{\ell\left(\lambda\right)}$, of \eqref{eqn:Hamiltonian} for an $n$-qubit system will be given by
\correction{
\begin{equation}
	\mathcal{H}(\lambda)\ket{\ell\left(\lambda\right)}=E_{\ell}\left(\lambda\right)\ket{\ell\left(\lambda\right)},
	\label{eqn:eigen}
\end{equation}
where $0\leq \ell \leq 2^{n}-1$ and $E_{0}<E_{1}<\ldots < E_{2^{n}-1}$}. In most AQC schemes, it is usually assumed that the interpolating parameter $\lambda(t)$ varies linearly with time and therefore the rate $d\lambda / dt=\rm{constant}$. 

	The critical point in the evolution of an adiabatic quantum computation is the minimum gap between the ground and first excited state, $\Delta_{\rm{min}}=\min_{0\leq\lambda\left(t\right)\leq 1}\left[E_{1}(\lambda)-E_{0}(\lambda)\right]$. \correction{For an AQC process described by a total Hamiltonian of the form \eqref{eqn:Hamiltonian}, $\Delta_{\rm{min}}$ will occur towards the end of the computation, when the levels become more densely packed as the relative strength of the perturbing $H_b$ is much less than $H_p$, as seen in \cite{Zagoskin2007,Wilson2010}.} In the vicinity of this avoided crossing, the chance of excitation through Landau-Zener-Stueckelberg (LZS) \cite{Shevchenko2010} tunnelling is at its maximum. The chance of excitation via the LZS mechanism increases monotonically with $d\lambda / dt$ as well as $\Delta_{\rm{min}}$. As $\Delta_{\rm{min}}$ is fixed by the choice of problem and bias Hamiltonians, the rate $d\lambda / dt$ must be chosen to be relatively slow in order to minimise the excitation probability. The optimum speed to cross the critical minimum gap in the presence of decoherence was derived in \cite{Ashhab2006}. This is the optimal speed for the entire computation, if it is performed at a constant rate. However, far away from the minimum gap, where the ground and first excited states are well separated, the computation could be run much faster than this optimal speed without significantly increasing the chance of excitation. This leads to the idea of designing time-optimal, non-linear interpolation paths, $\lambda(t)$, for AQC that take into consideration the critical points in the evolution. A number of approaches to realize this have been explored; \cite{Rezakhani2009,Rezakhani2010} propose a variational ansatz that allows an optimal path to be determined, and \cite{Avron2010} suggests that dephasing may pick out the optimal path under certain conditions. 
	
	In this paper we propose a feedback-control scheme for the interpolation rate based on the result of weak continuous measurements of the ground state curvature of a flux qubit AQC system. The aim is to allow the computation process to move quickly when the ground state is well separated from the excited states, but then gradually slow down as it approaches and traverses the critical minimum gap. \correction{After it has moved through the minimum gap, the rate of the computation can then be safely increased again.} We show that this type of control scheme can provide a significant improvement in the performance of existing solid state implementations of AQC without the need for additional on-chip hardware.
	
\section{Specifications for feedback-control of AQC}

	Feedback-control \cite{Wiseman2009} is a very common control scheme that has been applied to a wide range of applications. Essentially, this approach makes use of the results of continuous measurement of a dynamical system to continuously adjust its control parameters to regulate its output. 
	
	In terms of AQC, the  main process variable that we would like to control is the probability of successfully finding the system in the ground state of the problem Hamiltonian at the end of the evolution, when $\mathcal{H}(\lambda=0)=H_{p}$. This success probability is given by 
\begin{equation}
	P=\left|\braket{\psi(\lambda=0)}{0(\lambda=0)}\right|^2.
	\label{eqn:SuccessP}
\end{equation} 
The aim of a feedback-control scheme for AQC would be to increase $P$ without the need to also increase the total computation time $T$. The most easily-controlled parameter in \eqref{eqn:Hamiltonian} is the interpolation function $\lambda(t)$ and its corresponding rate of change $d\lambda / dt$. With relatively precise control of $d\lambda / dt$ throughout the evolution, it would be possible to move slowly through the critical minimum gap and then evolve quickly when the ground state is well separated from the rest of the spectrum. This type of scheme should in theory allow a given probability to be achieved with a shorter $T$ when compared to the linear interpolation of $\lambda(t)$, and therefore produce an improvement in the scaling of $P$ with $T$. 
	
	The key requirement needed to realise this scheme is a means of detecting when the system is moving through the critical minimum gap. Crucially, the measurement must have minimal back-action on the evolution of the system. It also needs to be continuous and allow readout on a timescale that is much shorter than the total computation of the AQC operation.
	
\begin{center}
\begin{figure}
	\includegraphics[width=\columnwidth]{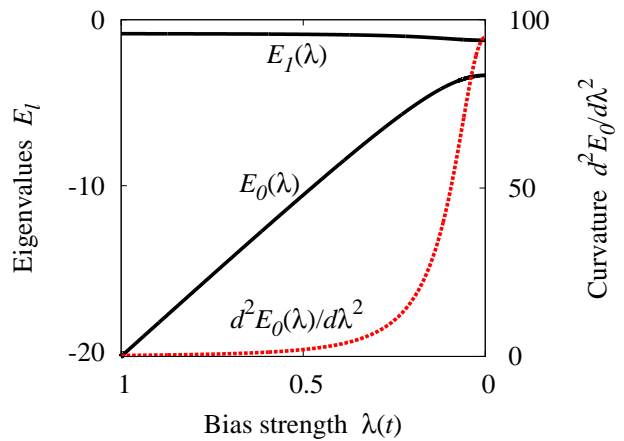}
	\caption{\correction{(Color online) Plot of the lowest two energy eigenvalues, $E_{0}(\lambda)$ and $E_{1}(\lambda)$ (black solid lines), and the curvature of the ground state, $d^{2}E_{0}(\lambda) / d\lambda^{2}$ (red dashed line), for a typical example of a 2-qubit adiabatic quantum computation.}}
	\label{fig:GapCurvature}
\end{figure}
\end{center}

	\correction{A method of weak continuous measurement, that has been succesfully applied to systems of superconducting flux qubits, is the impedance measurement technique (IMT), see \cite{Zagoskin2011} for a detailed review of IMT or e.g. \cite{Greenberg2002,Ilichev2004,vanderPloeg2007} for applications of it. In the IMT, a driven high-quality $LC$ tank circuit is inductively coupled to the flux qubit system. The coupling to the qubits will induce a frequency shift in the resonator  which is proportional to the curvature of the qubit system's state. This frequency shift can be monitored by measuring the impedance of the tank through the phase angle between the driving current and the tank voltage. In AQC, the system should remain in the instantaneous ground state throughout the evolution and therefore the results of the IMT will be proportional to the ground state curvature, $d^{2}E_{0}(\lambda) / d\lambda^{2}$. The ground state curvature is a particularly appropriate property of the system to monitor in a feedback-control loop, as we know that it will increase to its global maximum at the critical point $\Delta_{\rm{min}}$, as shown in the example of \figref{fig:GapCurvature}.}

	The back-action of the IMT measurement on the qubit system was studied in detail by Smirnov in \cite{Smirnov2003}. The conditions required to consider IMT as a weak continuous measurement of Rabi oscillations in a flux qubit were determined and it was found that the contribution to the decay rate of the qubit due to the back-action of the tank circuit is peaked when $\Omega_R=\omega_{LC}$, where $\Omega_R$ is the qubit Rabi frequency and $\omega_{LC}$ is the resonant frequency of the tank circuit. In this case the relevant frequency scale of the qubit system is $\Delta_{\rm{min}}/\hbar$ and we therefore require
\begin{equation}
	\frac{\Delta_{\rm{min}}}{\hbar}>\omega_{LC}+\gamma_{LC} \quad\text{or}\quad \frac{\Delta_{\rm{min}}}{\hbar}<\omega_{LC}-\gamma_{LC},
	\label{eqn:backActionInequality}
\end{equation}
where $\gamma_{LC}$ is the tank linewidth, in order to minimise the decoherence caused by the IMT measurement and allow it to be classified as weak continuous measurement. Requiring the two frequency scales to be out of resonance is a relatively easy requirement to meet practically in comparison to requiring them to be in resonance. This is because it still does not require exact knowledge of $\Delta_{\rm{min}}$, only an order of magnitude estimate. 

	Therefore, IMT is a promising potential candidate for the type of weak continuous measurement required to practically implement a feedback-control scheme for AQC, where we take
\begin{equation}
	\frac{d\lambda}{dt}=f\left(\frac{d^{2}E_{0}(\lambda)}{d\lambda^{2}}\right).
	\label{eqn:controllerRule}
\end{equation}
The IMT also has the advantage that the same $LC$ tank circuit can also be used for dispersive readout of the solution state, as in \cite{vanderPloeg2007}, which means that the proposed feedback-control scheme does not require any additional on-chip hardware.
	
	We envision two possible modes of operation for this scheme in experimental implementations: \textit{$\left(i\right)$} Offline control. Where one initial evolution of the system is performed with the weak continuous measurement to precompute $d^{2}E_{0}(\lambda) / d\lambda^{2}$. Then the actual AQC evolution is carried out using the resulting interpolation function as many times as necessary to collect the required statistics for the measurement of the solution state. \textit{$\left(ii\right)$} Online control. In this case the weak continuous measurement is performed throughout all of the AQC evolutions and the rate is adjusted actively through the feedback-control loop. In this case, the effects of the back-action of the measurement on the system will be a more important consideration.
	
\section{Theoretical model for feedback-controlled AQC}

	We assume that the inequality \eqref{eqn:backActionInequality} holds true and analyse the ideal case of a negligibly small back-action from the continuous IMT measurement on the qubit system. In order to simulate this system, we first need to accurately compute the ground state curvature as a function of $\lambda(t)$. The Pechukas-Yukawa equations of motion, \cite{Pechukas1983,Yukawa1985,Zagoskin2007,Wilson2010}, can be used to describe the dynamics of the energy eigenvalues of a quantum system with a Hamiltonian of the form \eqref{eqn:Hamiltonian};
\begin{align}
	\frac{\partial E_{\ell}}{\partial\lambda}=&v_{\ell}, \notag\\
	\frac{\partial v_{\ell}}{\partial\lambda}=&\sum_{k\neq \ell}\frac{2\left|l_{\ell k}\right|^{2}}{(E_{\ell}-E_{k})^{3}}, \label{eqn:PYSystem}\\
	\frac{\partial l_{\ell\jmath}}{\partial\lambda}=&\sum_{k\neq \ell,\jmath}l_{\ell k}l_{k\jmath}\left(\frac{1}{(E_{\ell}-E_{k})^2} - \frac{1}{(E_{\jmath}-E_{k})^2}\right), \notag
\end{align}
where $v_{\ell}=\bra{\ell}H_b\ket{\ell}$ and $l_{\ell\jmath}=(E_{\ell}-E_{\jmath})\bra{\ell}H_b\ket{\jmath}$. Then from \eqref{eqn:PYSystem}, the ground state curvature will be given by
\begin{equation}
	\frac{\partial^{2}E_{0}}{\partial\lambda^{2}}=\frac{-2\left|l_{01}\right|^2}{\left(E_{1}-E_{0}\right)^{3}}.
	\label{eqn:xcurvature}
\end{equation}

	The evolution of the state of the system, $\ket{\psi\left(\lambda\right)}$, will be described by the Schr\"{o}dinger equation 
\begin{equation}
	\frac{\partial}{\partial\lambda}\ket{\psi\left(\lambda\right)}=-i\left(\frac{d\lambda}{dt}\right)^{-1}\mathcal{H}\left(\lambda\right)\ket{\psi\left(\lambda\right)}.
	\label{eqn:schrodinger}
\end{equation}
The $\left(d\lambda / dt\right)^{-1}$ term in \eqref{eqn:schrodinger} describes the feedback controller rule and will, in general, be of the form \eqref{eqn:controllerRule}. The simplest type of feedback controller is a constant proportional gain controller and, as we wish the interpolation rate to decrease as the curvature increases near the critical minimum gap, we therefore take
\begin{equation}
	\frac{d\lambda}{dt}=\left(k\frac{d^{2}E_{0}}{d\lambda^{2}}\right)^{-1},
	\label{eqn:PController}
\end{equation}
where $k$ is the gain constant. The total computation time $T$ of a feedback-controlled AQC evolution will be given by
\begin{equation}
	T=\int^{1}_{0}\left(\frac{d\lambda}{dt}\right)^{-1}d\lambda.
	\label{eqn:compTime}
\end{equation}

\correction{\section{A prototypical example of feedback-controlled AQC}}

	We take the bias Hamiltonian $H_b$ to be a strong transverse field acting equally on all $n$ qubits with an easily constructed Hamiltonian of the form
\begin{equation}
	H_b=-Z\sum_{i=1}^{n}\sigma_{x}^{\left(i\right)},
	\label{eqn:Hb}
\end{equation}
where the constant $Z\gg1$ and $\sigma_a$ (with $a=x,y,z$) are the usual Pauli matrices. The ground state \ket{0_{b}} of $H_b$ is simply an equal superposition of all $2^n$ computational basis states,
\correction{
\begin{equation}
	\ket{0_{b}}=2^{-n/2}\sum_{j=0}^{2^{n}-1}\ket{j}.
\end{equation}} 
In the following calculations we take the $Z=10^{n/2}$.

	As a prototypical example of an AQC problem Hamiltonian we use a $H_p$ with randomly weighted coupling in the $z$-direction between all possible combinations of the $n$ qubits;
\begin{equation}
	H_p=\sum_{j=1}^{2^n-1}\epsilon_{j}\bigotimes_{i = 1}^n (\sigma_z)^{j_i},
	\label{eqn:Hp}
\end{equation}
where $j_i$ is the $i^{th}$ bit in the binary representation of $j$ and $\epsilon_j$ are the coupling strengths, which are selected from a Gaussian distribution with mean $\mu=0$ and standard deviation $\sigma=n^2$. This form of randomly selected problem Hamiltonian represents a reasonable approximation as it can be used to encode any finite computational optimisation problem with a proper choice of $\epsilon_j$.

	To investigate the efficacy of the proposed feedback-control scheme for AQC, we have solved equations \eqref{eqn:PYSystem} and \eqref{eqn:schrodinger} numerically for a range of different gain constants $k$ and system sizes $n$ for an ensemble of random $H_p$'s. We also solve \eqref{eqn:schrodinger} for a range of linear interpolations, where $d\lambda / dt=\rm{constant}$, so that we have a benchmark for comparison to the results for feedback-controlled AQC. \correction{In both cases the system is initially prepared in the ground state of the total Hamiltonian \eqref{eqn:Hamiltonian}, $\ket{0\left(\lambda=1\right)}$.}
	
\begin{center}
\begin{figure}
	\includegraphics[width=\columnwidth]{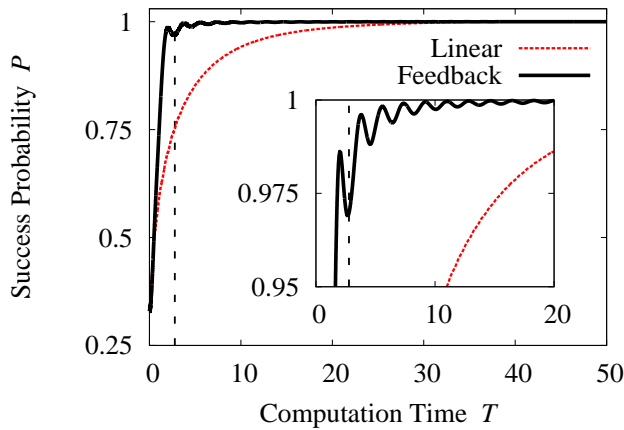}
	\caption{(Color online) Comparison of the scaling of the success probability $P$ versus the total computation time $T$ for linear and feedback-controlled interpolation for a typical example of a 2-qubit adiabatic quantum computation. \correction{The dashed black line indicates the adiabatic time, $T_{\rm{ad}}=2.775$, for the computation. The inset shows the non-adiabatic oscillations that arise for the feedback-controlled case around the adiabatic limit.}}
	\label{fig:2qbExample}
\end{figure}
\end{center}

	A comparison of the scaling of the success probability $P$ with the total computation time $T$ for a typical 2-qubit AQC operation with and without feedback-control is shown in \figref{fig:2qbExample}. In this case, the feedback-controlled interpolation causes $P$ to approach unity more rapidly than the standard linear interpolation scheme allows. This improvement can be interpreted in one of two ways; as an increase in $P$ at a given $T$, or conversely as a decrease in the $T$ required to achieve a given $P$. 

	\correction{Figure \ref{fig:2qbExample} also shows that the success probability, $P$, for the feedback-controlled interpolation oscillates slightly in the vicinity of the adiabatic time \cite{Amin2009},
\begin{equation}
	T_{\rm{ad}}=\max_{1\leq\jmath\leq 2^{n}-1}\left(\max_{0\leq\lambda\leq 1}\frac{\left|\bra{\jmath\left(\lambda\right)} \frac{d\mathcal{H}}{d\lambda} \ket{0\left(\lambda\right)}\right|}{\Delta_{\rm{min}}^{2}}\right).
\end{equation}
These non-adiabatic oscillations most likely arise from the system's non-linear trajectory through phase space, and despite them the feedback-control scheme still offers a significant improvement over linear interpolation. In the adiabatic limit, when $T\gg T_{\rm{ad}}$, feedback-controlled interpolation will still offer some improvement, but as $P$ will already be arbitrarily close to 1, this improvement will be negligible. The regime where the feedback-control method excels is the approximately adiabatic case, where $T \sim T_{\rm{ad}}$, as can be seen in \figref{fig:2qbExample}. In practical AQC devices we would like to achieve a relatively high success probability $P$ in the shortest time $T$ possible and it is the approximately adiabatic regime that offers the largest potential gains in this trade-off. This is because a small increase in $T$ can significantly increase $P$ and the proposed feedback-control scheme improves this situation even further.}

\begin{center}
	\begin{figure}
		\includegraphics[width=\columnwidth]{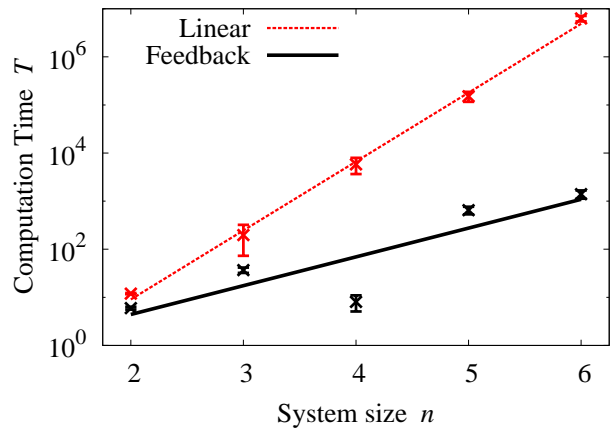}
		\caption{\correction{(Color online) Comparison of the average computation time $T$ required to achieve a  success probability of $P=90\%$ for linear and feedback-controlled interpolation as a function of the number $n$ of qubits. The polynomial lines of best fit give $T=O\!\left(n^{3.3}\right)$ for linear interpolation and $T=O\!\left(n^{1.4}\right)$ for the feedback-controlled case; an approximately quadratic speed-up. The results are averaged over 10,000 randomly selected problem Hamiltonians of the form \eqref{eqn:Hp} and the error bars denote the standard deviation of the distributions of $T$.}}
		\label{fig:TScaling}
	\end{figure}
\end{center}

	 \correction{A key issue for our proposed feedback-control approach is how well its performance scales with the system size. The dependence of the average computation time on the system size for linear and feedback-controlled interpolation is shown in \figref{fig:TScaling}. For linear interpolation we have $T=O\!\left(n^{3.3}\right)$, while for the feedback-controlled interpolation we find that $T=O\!\left(n^{1.4}\right)$. Therefore, in this case the proposed feedback-control method provides a polynomial speed-up of approximately $O\!\left(n^{2}\right)$.} This speed-up occurs because as $n$ increases the system's spectra become more densely packed and $\Delta_{\rm{min}}$ decreases  (as shown in \cite{Karimi2010}). This means that the speed at which $\Delta_{\rm{min}}$ is traversed becomes an even more important factor in determining whether the system is excited out of the ground state. Therefore, controlling the speed in proportion to the gap becomes an even more effective strategy. We note that it is impossible to infer exactly how the maximum $T$ will scale in the limit of large $n$ from these results though.

\begin{center}
\begin{figure}
	\includegraphics[width=\columnwidth]{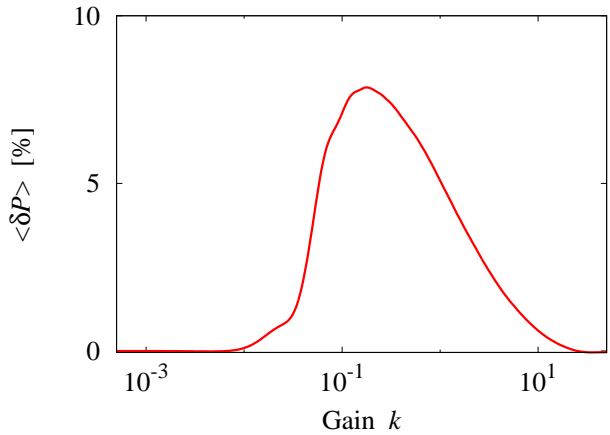}
	\caption{(Color online) The average percentage increase in the success probability $\left<\delta\! P\right>$ for feedback-controlled interpolation compared to linear interpolation as a function of the controller's gain $k$. \correction{These results are averaged over 10,000 different 2-qubit problem instances of the form \eqref{eqn:Hp}.}}
	\label{fig:PComparison}
\end{figure}
\end{center}

	Another consideration about the effectiveness of the proposed feedback-control scheme is whether it is possible to select an appropriate set of controller parameters, $k$, that will provide an improvement in the performance for a given $H_p$ without any \textit{a priori} knowledge of the evolution. We define the increase in success probability $\delta\! P$ as 
\begin{equation}
	\delta\! P=\frac{P_{\rm{fb}}-P_{\rm{lin}}}{P_{\rm{lin}}},
\end{equation}
where $P_{\rm{fb}}$ is the success probability for a feedback-controlled AQC operation with a given $k$ and total computation time $T$ and $P_{\rm{lin}}$ is the success probability for the same AQC operation with linear interpolation over the same time $T$. Figure \ref{fig:PComparison} shows that the average increase in success probability due to the feedback-control approach as a function of the controller gain is unimodal and that there is a clear optimal value of $k$. These results are averaged over an ensemble of random $H_p$'s which could represent a wide range of computational problems. Therefore, we can conclude that an appropriate value of $k$, which will usually provide an increase in $P$ for an arbitrary $H_p$, can be found. 

\section{Conclusions}	

	We have proposed a simple feedback-control scheme for solid-state implementations of AQC comprised of flux qubits. The scheme makes use of existing on-chip hardware to perform weak, continuous measurement of the ground state curvature. The interpolation speed is dynamically controlled as a function of the ground state curvature in order to maximise the success probability. The ground state curvature increases to a maximum at the minimum ground state gap, where the computation speed is reduced to avoid excitation. Then, as the levels spread back apart and the curvature decreases, the computation speed is increased with little risk of excitation. This approach can be interpreted as attempting to optimise the computation speed in a piece-wise fashion throughout the entire evolution, compared to calculating a single optimal speed based on the limiting minimum gap as in \cite{Ashhab2006}.
	
	As a theoretical proof of concept, we have shown that a simple proportional gain controller will provide a significant increase in performance when compared to the usual linear interpolation scheme for a class of generic problem Hamiltonians. \correction{Crucially, it was shown that this scheme can provide a polynomial speed-up, approximately $O\!\left(n^{2}\right)$, in the scaling of the computation time with system size}, although the exact nature of this scaling in the limit of large system sizes remains unclear. We have also shown that it should be possible to select appropriate controller parameters that will usually provide a significant increase in success probability for an arbitrary problem. \correction{We expect that the improvement gained from this scheme will be most noticeable in the approximately adiabatic regime, which is also where we would like practical AQC devices to operate, as it offers the best compromise between computation time and success probability.}  By tailoring the controller rule, e.g. using a more sophisticated non-linear controller rule, or by restricting ourselves to a more specific class of problems, it may be possible to achieve even greater improvements in performance using this feedback-control scheme.

\section*{Acknowledgements}
	R.W., A.Z., S.S. and M.E. acknowledge that this publication was made possible through the support of a grant from the John Templeton Foundation. F.N. is partially supported by the ARO, JSPS-RFBR contract No. 12-02-92100, Grant-in-Aid for Scientific Research (S), MEXT Kakenhi on Quantum Cybernetics, and the JSPS via its FIRST program.

\bibliography{feedbackdraft6refs}
\end{document}